# Consistency between observational and empirical data of the thermospheric $CO_2$ and NO power

C. A. VAROTSOS and M. N. EFSTATHIOU

*Department of Environmental Physics and Meteorology, National and Kapodistrian University of Athens, Athens, Greece University Campus Bldg. Phys. V, Athens 15784, GR*

**Abstract.** We explore the temporal evolution of the energy radiated by $CO_2$ and NO from the Earth's thermosphere on a global scale. This investigation is based on both observational and empirically derived data. Firstly, we analyze the daily power observations of $CO_2$ and NO obtained by the Sounding of the Atmosphere using Broadband Emission Radiometry (SABER) instrument on the NASA Thermosphere-Ionosphere-Mesosphere Energetics and Dynamics (TIMED) satellite throughout the period 2002 - 2016. Secondly, we perform the same analysis to the empirical daily power emitted by $CO_2$ and NO that were derived recently from the infrared energy budget of the thermosphere during 1947-2016. The tool employed for the analysis of the observational and the empirical datasets is the detrended fluctuation analysis, in order to investigate whether the power emitted by $CO_2$ and by NO in the thermosphere exhibits power-law behaviour. The results obtained from both the observational and empirical data do not support the establishment of the power-law behaviour. This result indicates that the empirically derived data exhibit the same intrinsic properties with the observational ones, thus enhancing their reliability.

## 1. Introduction

Although many papers discuss the moderate anthropogenic temperature changes expected in the lower atmosphere, other publications predict a severe cooling (of 10±15 K) in the upper stratosphere and mesosphere, in case of $CO_2$ doubling.

Last years a few papers have presented analysis of observations of the infrared radiative cooling by $CO_2$ and NO in the Earth's thermosphere during 2002 – 2009, provided by the Sounding of the Atmosphere using Broadband Emission Radiometry (SABER) instrument on the TIMED satellite [1]. The results obtained showed a large decrease in the cooling rates, fluxes, and power consistent with the declining phase of solar cycle. In addition, a substantial short-term variability in the infrared emissions throughout the entire mission duration has been observed.

Spectral analysis on the thermospheric $CO_2$ and NO daily global power from 2002 through 2006 showed a statistically significant 9-day periodicity in the infrared data but not in the solar data [2]. However, a strong 9-day periodicity was also detected in the time series of daily $A_p$ and $K_p$ geomagnetic indexes, revealing a link between the Sun and the infrared energy budget of the thermosphere.

An empirical model has recently been presented [3], where the $F_{10.7}$, *Ap*, and *Dst* indices were employed in the linear regression fitted to the time series of the thermospheric $CO_2$ and NO daily global power from 2002 through 2016, in order to develop the radiative cooling time series from 1947 to 2016. As it was derived, the total infrared energy radiated by the thermosphere, integrated over a solar cycle, seemed to be almost constant over the studied period, a fact that may assess the terrestrial context of the long-term record of solar-related indices.

The present study aims to detect long memory behaviour in the daily $CO_2$ and NO global power radiated from the Earth's thermosphere during 1947-2016, by using the above mentioned time series [3]. The feature of the long memory effect was earlier revealed in processes that are closely related to the total ozone content observations [4-18], the air temperature [19-21], the solar ultraviolet radiation [22-25] having strong impacts to the dynamics of the climate system [26-31].

## 2. Data and analysis

We herewith examine the temporal march of the daily global power (W) radiated by carbon dioxide ($CO_2$ at 15 $\mu$m) and by nitric oxide (NO at 5.3 $\mu$m) from the Earth's thermosphere between 100 km and 139 km altitude. $CO_2$ and NO daily power measurements (kindly provided by M. Mlynczak) cover 15 years from 2002 through 2016 (see [1-4] and have been taken by the Sounding of the Atmosphere using Broadband Emission Radiometry (SABER) instrument on the NASA Thermosphere-Ionosphere-Mesosphere Energetics and Dynamics (TIMED) satellite.

We also employ two more extended time series of thermospheric $CO_2$ and NO daily global power that cover the period from 1947 through 2016, developed recently [3]. Specifically, for the development of this this time series the following were used: the $F_{10.7}$, *Ap*, and *Dst* indices in linear regression fits to the above described data sets of $CO_2$ and NO power (2002-2016) to construct the infrared power emitted by NO and $CO_2$ back to 1947, which date is the beginning of the $F_{10.7}$ time series.

To study the scaling dynamics of all these time series, we use the DFA technique, which eliminates the noise of the non-stationarities that characterize the time series of $CO_2$ and NO daily power and permits the detection of intrinsic self-similarity [32-34]. The sequential steps of DFA are described in [34] and a detailed description is presented in [21].

The strong long-term trend and seasonality that characterize the time series of $CO_2$ and NO daily power are both removed (deseasonalization and detrending) by using the classical Wiener method [35] and the polynomial regression analysis, respectively.

In order to confirm the existence of long-range correlations in the time series of $CO_2$ and NO daily power, the autocorrelation function and the method of the local slopes of the fluctuation functions are also employed (i.e. the two criteria proposed in [36]).

## 3. Results and discussion

Our first step was to explore the temporal evolution of $CO_2$ and NO daily power, for the period 2002-2016. The initial time series are shown in Figure 1(a), while the corresponding root-mean-square fluctuation functions $F_d(\tau)$ of the DFA technique versus time scale $\tau$ (in days), are depicted in Figure 1(b).

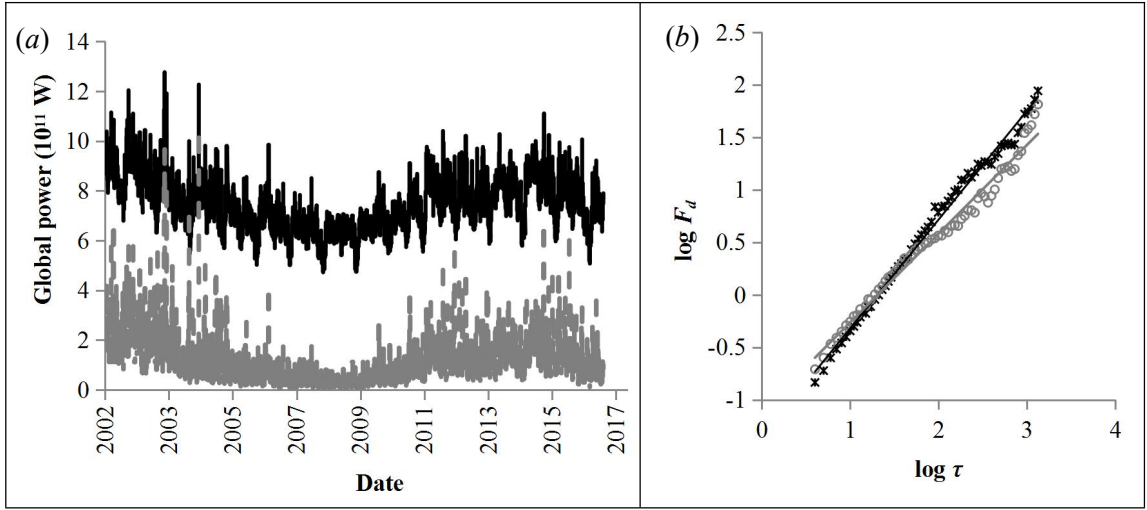

**Figure 1.** (a) Temporal march of the $CO_2$ (black line) and NO (gray line) daily power, during the period 2002-2016. (b) The corresponding root-mean-square fluctuation functions $F_d(\tau)$ of the DFA versus time scale $\tau$ (in days), in log-log plot and the respective best fit equations ($CO_2$: $y = 1.04x - 1.35$ with $R^2 = 0.99$, NO: $y = 0.84x - 1.11$ with $R^2 = 0.98$).

The derived DFA scaling exponent for the initial time series of $CO_2$ (NO) daily power was found $a = 1.04 \pm 0.01$ ($a = 0.84 \pm 0.02$), assuming therefore long-range persistence (of $1/f$ – type for the case of $CO_2$). However, to reliably establish power-law scaling and long-range correlations for the temporal march of $CO_2$ and NO daily power, the autocorrelation function and the local slopes $a$ vs $\log\tau$ should be investigated in relation to the criteria proposed in [36].

Specifically, Figure 2(a and b) presents the profiles of the power spectral density for the time series of $CO_2$ and NO daily power, showing therefore that the exponential decay could be rejected only for the case of $CO_2$, where power-law fit seems to give a little better coefficient of determination. Regarding the second criterion of [36], we fitted a straight line to $\log F_d(\tau)$ vs. $\log\tau$ (for both $CO_2$ and NO daily power) within a small window, shifting it successively over all calculated scales $\tau$. However, local slope $a$ vs. $\log\tau$ seemed to fluctuate without any interval of constancy, indicating thus that long-range correlations of power-law type could not be established either for $CO_2$ or for NO daily power (see Figure 3(a and b)).

Our second step was to re-apply the above described analysis on the detrended and deseasonalised time series of $CO_2$ and NO daily power, for the period 2002-2016 (see Figure 4(a and b). The derived DFA scaling exponent for the time series of $CO_2$ (NO) daily power was found $a = 0.78 \pm 0.01$ ($a = 0.63 \pm 0.01$), assuming again long-range persistence.

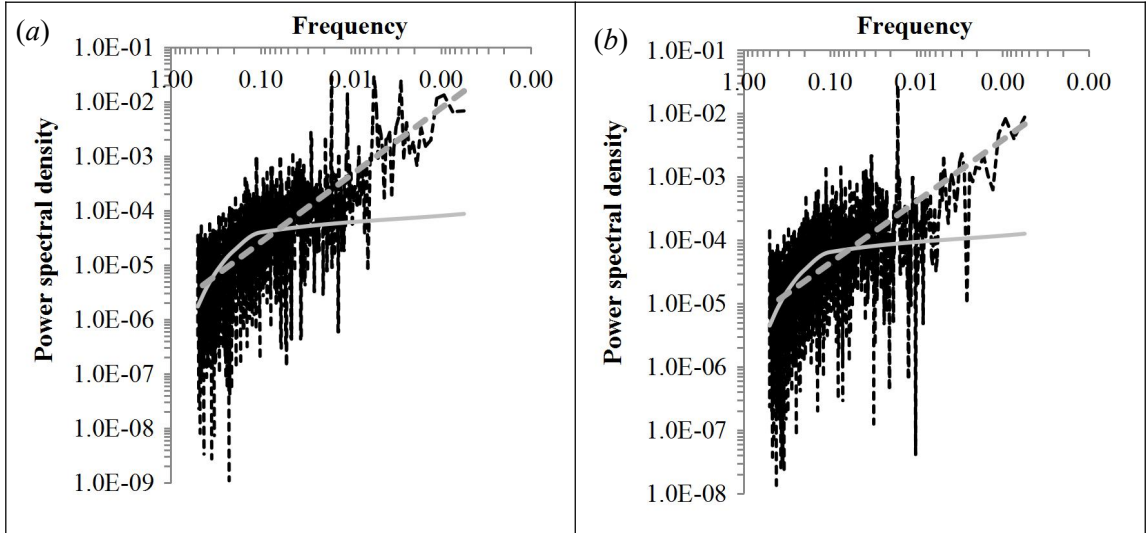

**Figure 2.** Power spectral density for the initial time series of (a) $CO_2$ daily power and (b) NO daily power (from 2002 to 2016), with the corresponding power-law (grey dashed line) and the exponential (grey solid line) fit ($CO_2$: $y = 1.54 \cdot 10^{-6} x^{-1.23}$ with $R^2 = 0.44$ and $y = 8.72 \cdot 10^{-5} e^{-7.84x}$ with $R^2 = 0.39$, NO: $y = 4.6 \cdot 10^{-6} x^{-0.97}$ with $R^2 = 0.34$, $y = 1.26 \cdot 10^{-4} e^{-6.68x}$ with $R^2 = 0.35$).

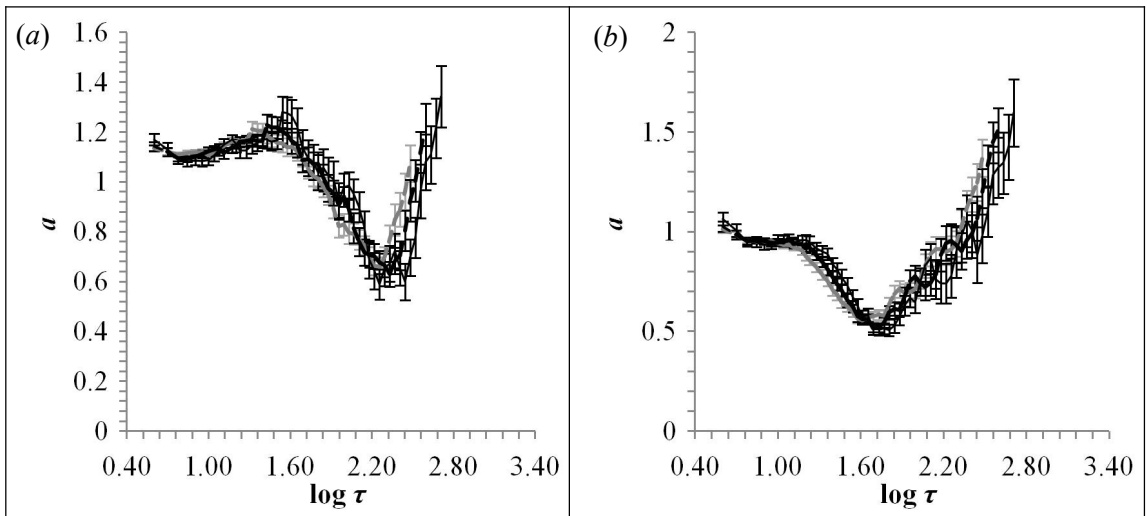

**Figure 3.** Local slopes of the $\log F_d(\tau)$ vs. $\log \tau$ (10-base logarithms) calculated within a window of 18 points (dashed grey line), of 12 points (solid thin black line) and of 15 points (dashed black line) for the initial time series of (a) $CO_2$ daily power and (b) NO daily power. The error bars indicate the corresponding $1.96 \cdot s_a$ – intervals of the slopes over all the considered scales.

Nevertheless, according to Figure 5(a,b) and Figure 6(a,b), the conditions of [36] seemed not to be satisfied, indicating once more that power-law scaling can not be established for the detrended and deseasonalised time series of $CO_2$ and NO daily power. It is worthy of note that only for the case of NO, there seems to be a constancy of the local slopes $a$ vs. $\log \tau$ in a small range, which is not however enough to confirm the long-range correlations.

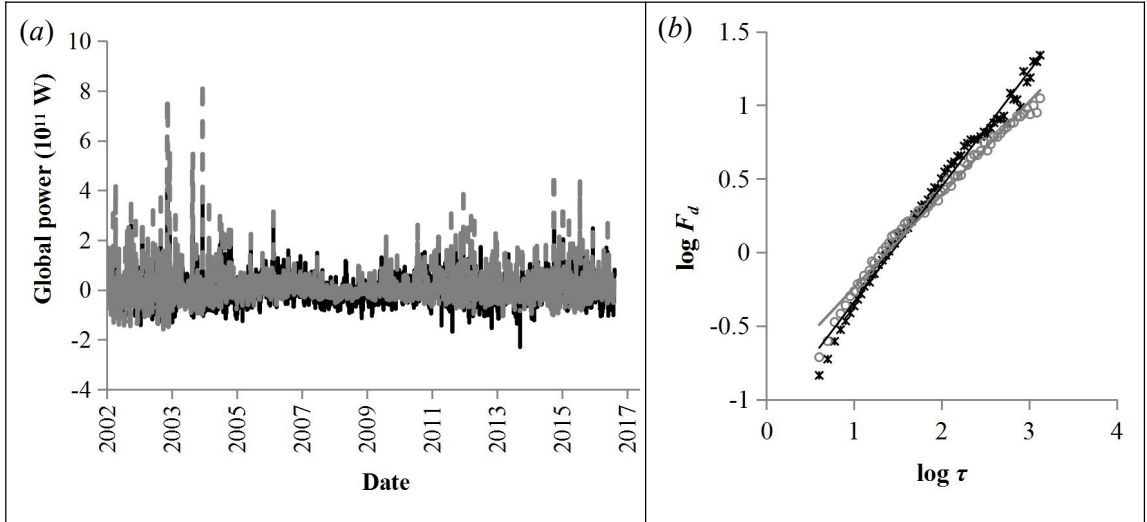

**Figure 4.** (a) Detrended and deseasonalised temporal march of the $CO_2$ (black line) and NO (gray line) daily power, during the period 2002-2016. (b) The corresponding root-mean-square fluctuation functions $F_d(\tau)$ of the DFA versus time scale $\tau$ (in days), in log-log plot and the respective best fit equations ($CO_2$: $y = 0.78x - 1.12$ with $R^2 = 0.99$, NO: $y = 0.63x - 0.87$ with $R^2 = 0.98$).

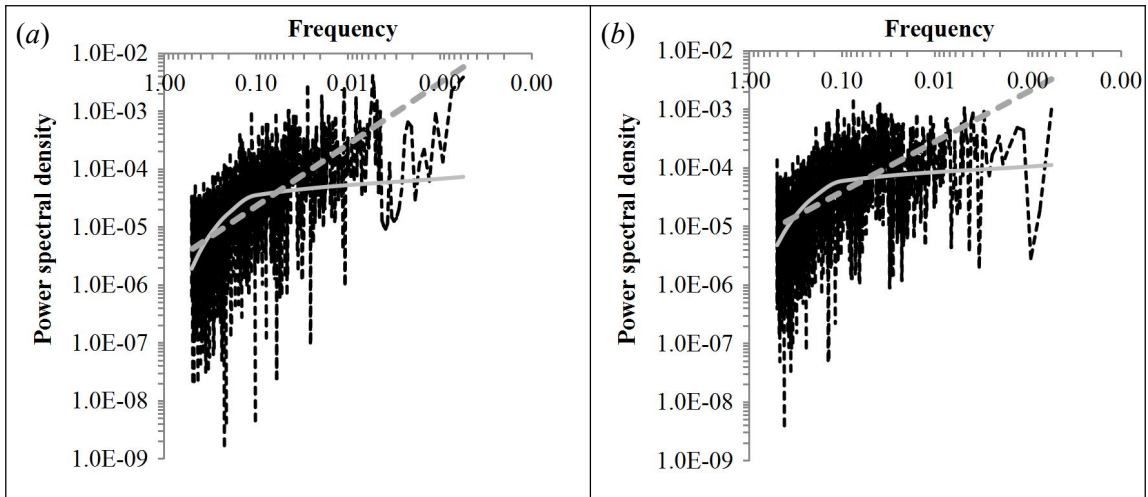

**Figure 5.** Power spectral density for the detrended and deseasonalised time series of (a) $CO_2$ daily power and (b) NO daily power (from 2002 to 2016), with the corresponding power-law (grey dashed line) and the exponential (grey solid line) fit ($CO_2$: $y = 1.95 \cdot 10^{-6} x^{-1.06}$ with $R^2 = 0.36$ and $y = 7.31 \cdot 10^{-5} e^{-7.31x}$ with $R^2 = 0.37$, NO: $y = 5.5 \cdot 10^{-6} x^{-0.85}$ with $R^2 = 0.28$, $y = 1.1 \cdot 10^{-4} e^{-6.31x}$ with $R^2 = 0.33$).

In the following, we studied the two other time series mentioned in the section 2, i.e. the radiated $CO_2$ and NO daily power, empirically derived from the infrared energy budget of the thermosphere from 1947 to 2016. The initial time series are shown in Figure 7(a), while the corresponding root-mean-square fluctuation functions $F_d(\tau)$ of the DFA technique versus time scale $\tau$ (in days), are depicted in Figure 7(b).

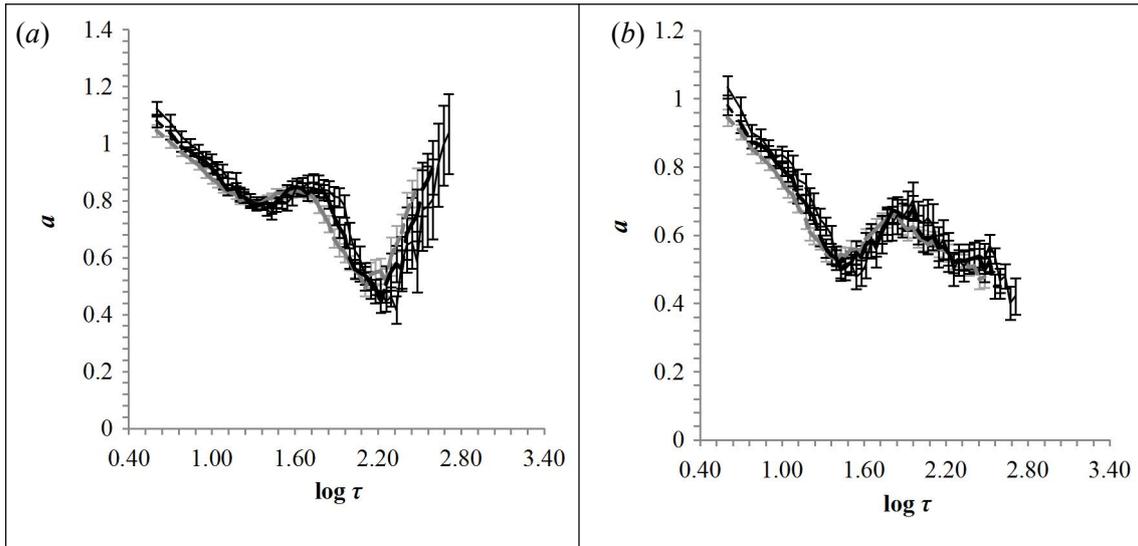

**Figure 6.** Local slopes of the $\log F_d(\tau)$ vs. $\log \tau$ (10-base logarithms) calculated within a window of 18 points (dashed grey line), of 12 points (solid thin black line) and of 15 points (dashed black line) for the detrended and deseasonalised time series of (a) $CO_2$ daily power and (b) NO daily power. The error bars indicate the corresponding $1.96 \cdot s_a$ – intervals of the slopes over all the considered scales.

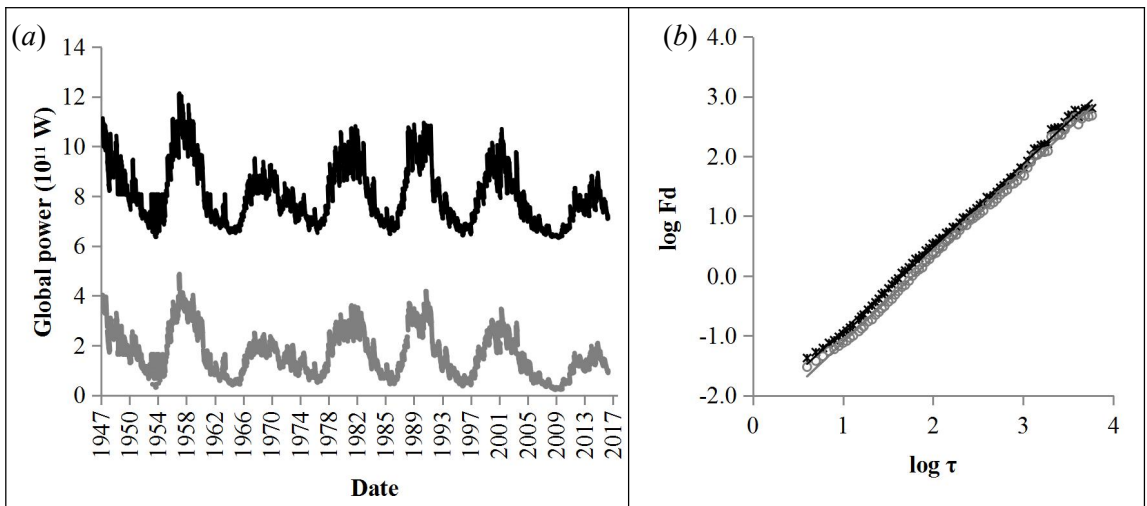

**Figure 7.** (a) The empirically derived infrared energy budget of the thermosphere from 1947 to 2016. The initial temporal march of the $CO_2$ (black line) and NO (grey line) daily power. (b) The corresponding root-mean-square fluctuation functions $F_d(\tau)$ of the DFA versus time scale $\tau$ (in days), in log-log plot and the respective best fit equations ($CO_2$: $y = 1.40x - 2.31$ with $R^2 = 0.998$, NO: $y = 1.43x - 2.54$ with $R^2 = 0.998$).

The derived DFA scaling exponent for the initial time series of $CO_2$ (NO) daily power (from 1947 to 2016) was found $a = 1.4 \pm 0.01$ ($a = 1.43 \pm 0.01$), assuming again long-range persistence. The algebraically (power law) fit gave better results than the exponential one for the power spectral density of both data sets (Figure 8(a,b)). Also, the local slope $a$ vs. $\log \tau$ (in 3 different window sizes of 24, 22 and 15 points) seemed to fluctuate again without any interval of constancy for both data sets (Figure 9(a,b)).

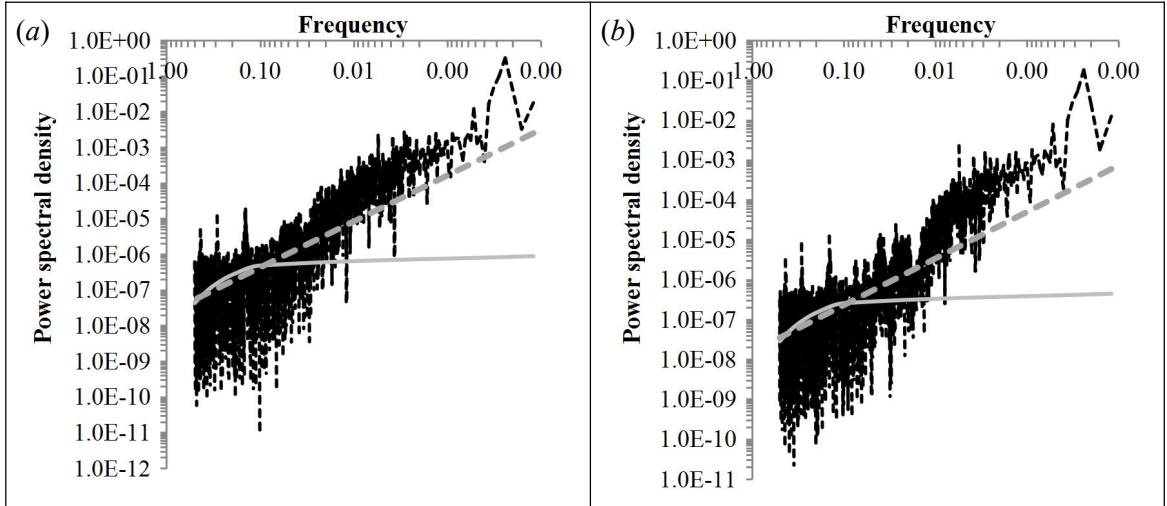

**Figure 8.** Power spectral density for the initial time series of (a) $CO_2$ daily power and (b) NO daily power (from 1947 to 2016), with the corresponding power-law (grey dashed line) and the exponential (grey solid line) fit ($CO_2$: $y = 2.19 \cdot 10^{-8} x^{-1.29}$ with $R^2 = 0.42$ and $y = 8.86 \cdot 10^{-7} e^{-6.07x}$ with $R^2 = 0.20$, NO: $y = 1.55 \cdot 10^{-8} x^{-1.17}$ with $R^2 = 0.39$, $y = 4.43 \cdot 10^{-7} e^{-5.5x}$ with $R^2 = 0.18$).

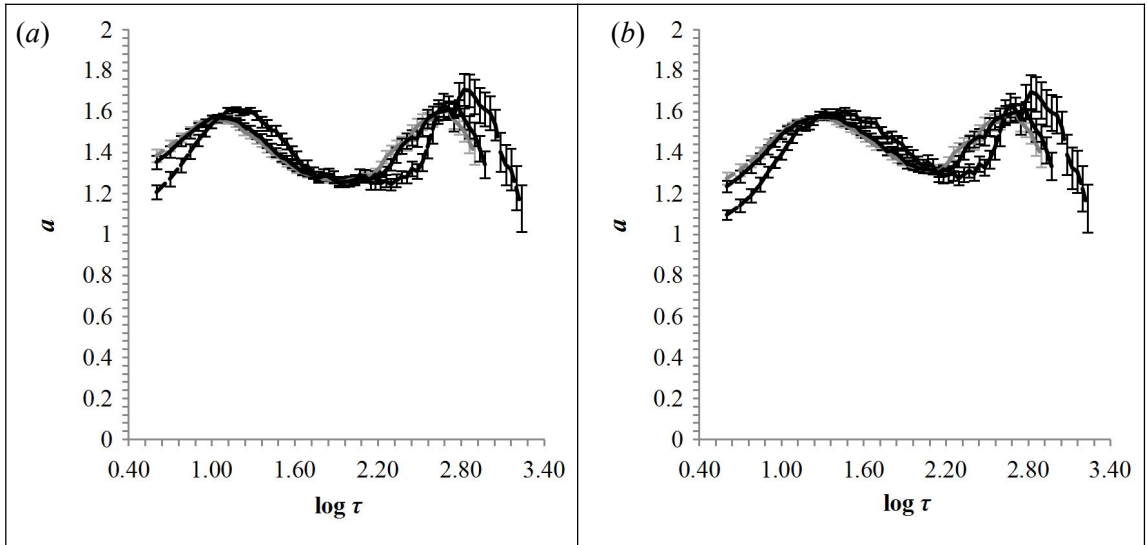

**Figure 9.** Local slopes of the $\log F_d(\tau)$ vs. $\log \tau$ (10-base logarithms) calculated within a window of 24 points (dashed grey line), of 22 points (solid thin black line) and of 15 points (dashed black line) for the initial time series of (a) $CO_2$ daily power and (b) NO daily power (from 1947 to 2016). The error bars indicate the corresponding $1.96 \cdot s_a$ – intervals of the slopes over all the considered scales.

Our final step was to re-apply the above described analysis on the detrended and deseasonalised time series of $CO_2$ and NO daily power, for the period 1947-2016. Table 1 illustrates the best fit equations of the DFA technique and the results of both criteria proposed in [36]. Thus, it seemed again that long-range correlations of power-law type are not established either for $CO_2$ or for NO daily power time series (empirically derived from the infrared energy budget of the thermosphere from 1947 to 2016).

Table 1. DFA, power spectral density and local slopes applied on the detrended and deseasonalised time series of $CO_2$ and NO daily power, for the period 1947-2016.

| | Best fit equations of the DFA method | Exponential and Power-law fit on the power spectral density | Local slope $a$ vs $\log\tau$ |
|---|---|---|---|
| $CO_2$ | $y = 1.39x - 2.31$ with $R^2 = 0.997$ crossover at 5 months $y = 0.38x - 0.08$ with $R^2 = 0.94$ | $y = 2.63 \cdot 10^{-8} \cdot x^{-1.18}$ with $R^2 = 0.38$ and $y = 8.47 \cdot 10^{-7} \cdot e^{-5.91x}$ with $R^2 = 0.2$ | *local slope a decreases without any interval of constancy* |
| NO | $y = 1.22x - 2.34$ with $R^2 = 0.996$ crossover at 5 months $y = 0.26x - 2.34$ with $R^2 = 0.96$ | $y = 2.25 \cdot 10^{-8} \cdot x^{-0.87}$ with $R^2 = 0.26$ and $y = 2.97 \cdot 10^{-7} \cdot e^{-4.44x}$ with $R^2 = 0.14$ | *local slope a decreases without any interval of constancy* |

## 4. Conclusions

1) In the present analysis we attempted to explore the temporal evolution of $CO_2$ and NO daily power, for the period 2002-2016. Although the derived DFA scaling exponent for the initial time series of both parameters assumed persistent behavior, power-law scaling and long range correlations were not established either for $CO_2$ or for NO daily power.
2) Similar results were extracted studying the scaling dynamics of the detrended and deseasonalised time series of $CO_2$ and NO daily power, for the period 2002-2016.
3) We also investigated the temporal march of the radiated $CO_2$ and NO daily power (empirically derived from the infrared energy budget of the thermosphere from 1947 to 2016). Although the derived DFA scaling exponent for the initial time series of both parameters assumed again persistent behavior, power-law scaling and long range correlations were established either for $CO_2$ or for NO daily power.
4) Similar results were extracted studying the scaling dynamics of the detrended and deseasonalised time series of $CO_2$ and NO daily power, for the period 1947-2016.

The detection of the scaling properties of $CO_2$ and NO power time series may lead to more reliable prediction of the expected temperature decrease in the upper atmosphere, discussed in the recent scientific literature, which can potentially lead to substantial changes in the structure and composition of the atmosphere [37-41].

## References

[1] Mlynczak, M. G., Hunt, L. A., Marshall, B. T., Martin-Torres, F. J., Mertens, C. J., Russell III, J. M., Remsberg, E. E., López-Puertas, M., Picard, R., Winick, J., Wintersteiner, P., Thompson, R. E. & Gordley, L. L. (2010). Observations of infrared radiative cooling in the thermosphere on daily to multiyear timescales from the TIMED/SABER instrument. Journal of Geophysical Research, 115, A03309, doi:10.1029/2009JA014713

[2] Mlynczak, M. G., Martin-Torres, F. J., Mertens, C. J., Marshall, B. T., Thompson, R. E., Kozyra, J. U., Remsberg, E. E., Gordley, L. L., Russell III, J. M. & Woods, T. (2008). Solar-terrestrial coupling evidenced by periodic behavior in geomagnetic indexes and the infrared energy budget of the thermosphere. Geophysical Research Letters 35 (5): L05808, doi: 10.1029/2007GL032620.

[3] Mlynczak, M. G., Hunt, L. A., Russell III, J. M., Marshall, B. T., Mertens, C. J. & Thompson, R. E. (2016). The global infrared energy budget of the thermosphere from 1947 to 2016 and implications for solar variability. Geophysical Research Letters 43: 11.934-11.940, doi:10.1002/2016GL070965.

[4] Varotsos, C.A. & Cracknell, A. P. (1993). Ozone depletion over Greece as deduced from Nimbus-7 TOMS measurements. International Journal of Remote Sensing, 14(11), 2053-2059.

[5] Varotsos, C.A. & Cracknell, A.P. (1994). Three years of total ozone measurements over Athens obtained using the remote sensing technique of a Dobson spectrophotometer. International Journal of Remote Sensing, 15(7), 1519-1524.

[6] Cracknell, A.P., & Varotsos, C.A. (1994) Ozone depletion over Scotland as derived from Nimbus-7 TOMS measurements. International Journal of Remote Sensing, 15(13), 2659-2668.

[7] Cracknell, A.P., & Varotsos, C.A. (1995) The present status of the total ozone depletion over Greece and Scotland: a comparison between Mediterranean and more northerly latitudes. International Journal of Remote Sensing, 16(10), 1751-1763.

[8] Kondratyev, K.Y., Varotsos, C.A., & Cracknell, A.P. (1994). Total ozone amount trend at St Petersburg as deduced from Nimbus-7 TOMS observations.International Journal of Remote Sensing, 15(13), 2669-2677.

[9] Reid, S.J., Rex, M., Von Der Gathen, P., Fløisand, I., Stordal, F., Carver, G.D., ... & Braathen, G. (1998). A Study of Ozone Laminae Using Diabatic Trajectories, Contour Advection and Photochemical Trajectory Model Simulations. Journal of Atmospheric Chemistry, 30(1), 187-207.

[10] Cracknell, A.P., & Varotsos, C.A. (2007). Editorial and cover: Fifty years after the first artificial satellite: from sputnik 1 to envisat.

[11] Varotsos, C., Kalabokas, P., & Chronopoulos, G. (1994). Association of the laminated vertical ozone structure with the lower-stratospheric circulation. Journal of Applied Meteorology, 33(4), 473-476, 1994.

[12] Gernandt, H., Goersdorf, U., Claude, H., & Varotsos, C.A. (1995). Possible impact of polar stratospheric processes on mid-latitude vertical ozone distributions. International Journal of Remote Sensing, 16(10), 1839-1850.

[13] Varotsos, C.A. & Tzanis, C. (2012). A new tool for the study of the ozone hole dynamics over Antarctica, Atmos. Environ., 47, 428–434.

[14] Varotsos, C. (2002). The southern hemisphere ozone hole split in 2002. Environmental Science and Pollution Research, 9(6), 375-376.

[15] Efstathiou, M. N., Varotsos, C. A., Singh, R. P., Cracknell, A. P., & Tzanis, C. (2003). On the longitude dependence of total ozone trends over middle-latitudes. International Journal of Remote Sensing, 24(6), 1361-1367.

[16] Reid, S.J., Vaughan, G., Mitchell, N.J., Prichard, I.T., Smit, H.J., Jorgensen, T.S., ... & De Backer, H. (1994). Distribution of ozone laminae during EASOE and the possible influence of inertia‐gravity waves. Geophysical Research Letters, 21(13), 1479-1482.

[17] Varotsos, C.A., Kondratyev, K.Y. & Cracknell, A.P. (2000). New evidence for ozone depletion over Athens, Greece. International Journal of Remote Sensing, 21(15), 2951-2955.

[18] Varotsos, C. & Cartalis, C. (1991). Re-evaluation of surface ozone over Athens, Greece, for the period 1901–1940. Atmospheric Research, 26(4), 303-310.

[19] Efstathiou, M.N., Tzanis, C., Cracknell, A.P. & Varotsos, C.A (2011). New features of land and sea surface temperature anomalies. International journal of Remote Sensing, 32(11), 3231-3238.

[20] Varotsos, C., Efstathiou, M. & Tzanis, C. (2009). Scaling behaviour of the global tropopause. Atmospheric Chemistry and Physics, 9(2), 677-683.

[21] Efstathiou, M.N. & Varotsos, C.A. (2010). On the altitude dependence of the temperature scaling behaviour at the global troposphere. International Journal of Remote Sensing 31, 343-349.

[22] Varotsos, C.A. & Cracknell, A.P. (2004). New features observed in the 11-year solar cycle. International Journal of Remote Sensing, 25(11), 2141-2157.

[23] Feretis, E., Theodorakopoulos, P., Varotsos, C., Efstathiou, M., Tzanis, C., Xirou, T., ... & Aggelou, M. (2002). On the plausible association between environmental conditions and human eye damage. Environmental Science and Pollution Research, 9(3), 163-165.

[24] Varotsos, C., Kondratyev, K. Y., & Katsikis, S. (1995). On the relationship between total ozone and solar ultraviolet radiation at St. Petersburg, Russia. Geophysical Research Letters, 22(24), 3481-3484.

[25] Efstathiou, M., Varotsos, C., & Kondratyev, K.Y. (1998). An estimation of the surface solar ultraviolet irradiance during an extreme total ozone minimum. Meteorology and Atmospheric Physics, 68(3), 171-176.

[26] Varotsos, C.A., Tzanis, C.G., & Sarlis, N.V. (2016). On the progress of the 2015–2016 El Niño event. Atmospheric Chemistry and Physics, 16(4), 2007-2011.

[27] Kondratyev, K. & Varotsos, C. (1995). Atmospheric greenhouse effect in the context of global climate change. Il Nuovo Cimento C, 18(2), 123-151.

[28] Cracknell, A.P. & Varotsos, C.A. (2011). New aspects of global climate-dynamics research and remote sensing. International Journal of Remote Sensing 32(3), 579-600.

[29] Krapivin, V.F. & Varotsos, C.A. (2016). Modelling the CO2 atmosphere-ocean flux in the upwelling zones using radiative transfer tools. Journal of Atmospheric Solar-Terrestrial. Physics, 150-151, 47-54.

[30] Krapivin, V.F., Varotsos, C.A. & Soldatov, V.Yu. (2015). New Ecoinformatics Tools in Environmental Science: Applications and Decision-making. Springer, London, U.K., 903 pp.

[31] Krapivin, V. F., & Varotsos, C. A. (2008). Biogeochemical cycles in globalization and sustainable development. Springer/Praxis, Chichester.

[32] Peng, C.-K., Buldyrev, S. V., Havlin, S., Simons, M., Stanley, H. E. & Goldberger, A. L. (1994). Mosaic organization of DNA nucleotides. Physical Review E 49, 1685–1689.

[33] Weber, R. O. & Talkner, P. (2001). Spectra and correlations of climate data from days to decades. Journal of Geophysical Research 106, 20131–20144.

[34] Varotsos, C. (2005). Modern computational techniques for environmental data; application to the global ozone layer. Computational Science–ICCS 2005, 43-68.

[35] Wiener N. (1950). Extrapolation, Interpolation and Smoothing of Stationary Time Series. *MIT Technology Press and John Wiley and Sons*, New York.

[36] Maraun, D., Rust, H. W. &. Timmer, J. (2004). Tempting long-memory – on the interpretation of DFA results. Nonlinear Processes in Geophysics 11, 495–503.

[37] Tzanis, C., Varotsos, C., & Viras, L.(2008). Impacts of the solar eclipse of 29 March 2006 on the surface ozone concentration, the solar ultraviolet radiation and the meteorological parameters at Athens, Greece. Atmospheric chemistry and physics, 8(2), 425-430..

[38] Varotsos, C., Tzanis, C., & Cracknell, A. (2009). The enhanced deterioration of the cultural heritage monuments due to air pollution. Environmental Science and Pollution Research, 16(5), 590-592.

[39] Tidblad, J., Kucera, V., Ferm, M., Kreislova, K., Brüggerhoff, S., Doytchinov, S., ... & Roots, O. (2012). Effects of air pollution on materials and cultural heritage: ICP materials celebrates 25 years of research. International Journal of Corrosion.

[40] Tzanis, C., Varotsos, C., Ferm, M., Christodoulakis, J., Assimakopoulos, M. N., & Efthymiou, C. (2009). Nitric acid and particulate matter measurements at Athens, Greece, in connection with corrosion studies. Atmospheric Chemistry and Physics, 9(21), 8309-8316.

[41] Tzanis, C., Varotsos, C., Christodoulakis, J., Tidblad, J., Ferm, M., Ionescu, A., ... & Kreislova, K. (2011). On the corrosion and soiling effects on materials by air pollution in Athens, Greece. Atmospheric Chemistry and Physics, 11(23), 12039-12048.